\documentclass[sn-chicago]{sn-jnl}

\usepackage{graphicx}%
\usepackage{multirow}%
\usepackage{amsmath,amssymb,amsfonts}%
\usepackage{amsthm}%
\usepackage{mathrsfs}%
\usepackage[title]{appendix}%
\usepackage{xcolor}%
\usepackage{textcomp}%
\usepackage{manyfoot}%
\usepackage{booktabs}%
\usepackage{algorithm}%
\usepackage{algorithmicx}%
\usepackage{algpseudocode}%
\usepackage{listings}%
\usepackage{placeins}

\theoremstyle{thmstyleone}%
\theoremstyle{thmstyletwo}%

\theoremstyle{thmstylethree}%

\begin{document}

\title[Article Title]{Statistical Challenges in Analyzing Migrant Backgrounds Among University Students: a Case Study from Italy}


\author*[1]{\fnm{Lorenzo} \sur{Giammei}}\email{lorenzo.giammei@cnr.it}

\author[2]{\fnm{Laura} \sur{Terzera}}\email{laura.terzera@unimib.it}

\author[2]{\fnm{Fulvia} \sur{Mecatti}}\email{fulvia.mecatti@unimib.it}

\affil*[1]{ \orgname{Italian National Research Council (CNR)}, \orgaddress{\street{Piazzale Aldo Moro, 7}, \city{Rome}, \postcode{00185}, \country{Italy}}}

\affil[2]{ \orgname{ University of Milano-Bicocca}, \orgaddress{\street{Piazza dell'Ateneo Nuovo, 1}, \city{Milan}, \postcode{20126}, \country{Italy}}}

\abstract{The methodological issues and statistical complexities of analyzing university students with migrant backgrounds is explored, focusing on Italian data from the University of Milano-Bicocca. With the increasing size of migrant populations and the growth of the second and middle generations, the need has risen for deeper knowledge of the various strata of this population, including university students with migrant backgrounds. This presents challenges due to inconsistent recording in university datasets. By leveraging both administrative records and an original targeted survey  we propose a methodology  to fully identify the study population of students with migrant histories, and to distinguish relevant subpopulations within it such as  second-generation born in Italy. Traditional logistic regression and  machine learning random forest models are used and compared  to predict migrant status. The primary contribution lies in creating an expanded administrative dataset enriched with indicators of students' migrant backgrounds and status. The expanded dataset provides a critical foundation for analyzing the characteristics of students with migration histories across all variables routinely registered in the administrative data set. Additionally, findings highlight the presence of selection bias in the targeted survey data, underscoring the need of further research.}




\maketitle

\section{Introduction}

The presence of foreigners in Italy has grown over time, and the reunification and family formation phase intensified at the end of the first decade of the new millennium. The Italian Statistical Agency (ISTAT)  found that on 1 January 2018 in Italy there were $1.316$ million second-generation minors, foreigners or Italians by acquisition, $75\%$ of whom were born in Italy \citep{Conti2020}. The largest proportion of those latter 
are mainly found in communities with longer immigration history, leading to both family reunification and family formation over time. Family rootedness has gradually brought the educational issue of immigrant children to the fore, and studies of pre-university performance and school choice are widespread in the literature. The main findings highlight, on the one hand, a tendency towards ghetto-isation in VET (vocational and technical) schools, with lower academic performance than the native populations as well as, on the other hand, higher drop-out rates \citep{Strozza2015}. With the growth of the second (and middle) generation, the study of the various school years has thus expanded to recently include the study of university students with a foreign background \citep{Bertozzi2018,Giudici2021,Giudici2022}.
As Table \ref{tab:foreign_students} shows, the number of international students (i.e. those with foreign citizenship and a foreign qualification) enrolled in Italian universities has risen sharply over the last ten academic years, and this increase has halted the (mainly demographic) steady decline in university enrolments in Italy \citep{Giudici2022,Giudici2023}

\begingroup
\setlength{\tabcolsep}{6pt}
\renewcommand{\arraystretch}{1.2}
\begin{table}[ht]
\centering
\scriptsize
\caption{International students enrolled in Italian three-year, master's or single-cycle programmes, basic index numbers (N.I. base academic year 2013/14) and \% of total enrolment \\ Source: Elaboration on MUR (Ministry of University and research) data}
\begin{tabular}{lccccccccccc}
\toprule
 & 13/14 & 14/15 & 15/16 & 16/17 & 17/18 & 18/19 & 19/20 & 20/21 & 21/22 & 22/23 \\ 
 \midrule
N.I. &  & 99.5 & 102.7 & 110.3 & 118.9 & 120.0 & 130.0 & 139.9 & 149.8 & 165.2 \\ 
\% & 4.8 & 4.8 & 4.8 & 4.9 & 5.3 & 5.2 & 5.3 & 5.4 & 5.8 & 6.4 \\ 
\bottomrule \\
\end{tabular}
\label{tab:foreign_students}
\end{table}
\endgroup

Nonetheless, the literature on the topic shows that the proportion of foreign citizens with a foreign pre-univeristy diploma remains constant because Italy is unattractive to students, while the number of children of immigrants is increasing \citep{Aiello2020}. This distinction is important because, although the university careers of students with a migrant background are slower and associated with higher drop-out rates, when international students are distinguished, it can be observed that the second generation (immigrants' children born in Italy) has a higher percentage of graduates and a lower drop-out rate \citep{Giudici2022}.
Literature focuses on two sub-groups of the population with a foreign background: international students, and those who instead obtained a degree in Italy 
while having
foreign citizenship or a place of birth other than Italy. Ultimately, it is these categories that are the main focus of the studies since the second generation (born in Italy) with Italian citizenship (because they are the children of naturalized immigrants or were naturalized themselves at the age of 18) cannot be statistically distinguished from Italians based on the official sources of the National Student Register nor on the archives of individual universities \citep{Giudici2021}. Similarly, the children of mixed couples, i.e. of an Italian and a foreigner parents, cannot be identified  if they were born in Italy, as they acquire Italian citizenship from birth.
This reality is most intense in the part of the country where immigration has been most intense and rooted over time, namely the northern regions and especially Lombardy. Therefore, Milan appears particularly interesting among the Italian university realities, not only as a pole of attraction for foreign students but also as a basin that gathers the children of immigrants, who are becoming increasingly numerous in the area.

The University of Milano-Bicocca (hereafter Unimib), through its Equality Committee (CUG) and other institutional bodies involved with international relations, has paid particular attention to the population of students with a migrant background in recent years, through targeted initiatives and support policies. 
In particular, in 2019 Unimib signed the Memorandum of Understanding of Inclusive Universities, promoted by the UN Refugee Agency (UNHCR) in order to encourage and guarantee equal opportunities for people with  migrant backgrounds to access quality learning opportunities at different levels, and especially in higher education. Examples of initiatives on the table were: the University Corridors for Refugees - UNICORE programme, coordinated by UNHCR, that aims to increase opportunities for refugees by offering scholarships to continue their higher education in Italy; and the project Mentorship in Italian Universities - Youth-to-Youth Support for the Integration of Students from Diverse Backgrounds,  implemented by the International Organisation for Migration (IOM) in coordination with the Italian Ministry of the Interior - Department of Civil Liberties and Immigration.
However, at present university registries do not include  students' migrant backgrounds and only cover a small part of the population that could generally be included in this category, e.g. international students such as Erasmus+ students, students who have obtained a degree abroad, and who have foreign nationality. This leaves statistically invisible the strata of students with different types of migration trajectories, including second generations born in Italy,  first generation who arrived in our country as pre-schoolers and acquired citizenship, children of mixed couples. The fact that the study population cannot be fully identified within university  records leads to incomplete data for statistical analysis.
This lack of data has motivated the implementation of a sample survey to obtain original information beyond those routinely recorded in university archives and, with a focus on strata of migrant students that are otherwise not identifiable. This survey was the basis for the UNI4ALL research project, sponsored by Unimib Equality Committee (CUG),  with the following areas of investigation: Origins, Family and home, Languages known and spoken, Schooling and university paths, Knowledge of university services for students with a migration background, Social networks, Discrimination within the school and university system.
However, the lack of a proper sampling frame for the study population of student with different type of migrant background, prevents the design of an appropriate survey sample, and  poses both methodological issues and statistical challenges, which are the main focus of the present paper.

The rest of the paper is organized as follows.
Section \ref{sec:data} provides a detailed description of available data for the purposes of UNI4ALL project, focusing on the main features of the Milano-Bicocca administrative records and the targeted survey, and addressing the emerging statistical challenges. Section \ref{sec:method} introduces a statistical strategy to comprehensively map the target population of Unimib students with migrant backgrounds, for integration into the administrative dataset. This strategy relies on a system of indicators to identify membership and typology of migrant background, along with models to predict hidden subgroups arising from gaps in administrative data. Results and concluding remarks are presented and discussed in Section \ref{sec:resdisc}.

\section{ Data and Statistical Challenges}
\label{sec:data}

\noindent This section provides details on the datasets available for our case study to analyze the migrant backgrounds of university students,  and discusses the statistical challenges that motivate the methodology proposed in this paper. We begin by defining the target population, followed by a description of the two main data sources: complete-coverage administrative archives and the UNI4ALL targeted sample survey.

\subsection{Target population}
\label{subsec:TP}
According to UNI4ALL scope and objectives, the target (or study) population with a migrant background is defined by all students enrolled at UniMiB that meet at least one of the following three criteria:

\begin{enumerate}
    \item not Italian (foreign, henceforth) citizenship;
   \item foreign country of birth; 
  \item  at least one parent with foreign birth-nationality.
\end{enumerate}
This definition will be used throughout the paper and constitutes a fundamental assumption to highlight the emerging statistical challenges and motivate the adopted statistic methodology.

\vspace{1cm}

\subsection{Administrative data}   
\label{subsec:AdmData}

For the purposes of the UNI4ALL project, Unimib provided administrative archives that offer comprehensive access to data recorded by university offices on all enrolled students. This longitudinal administrative dataset spans the period from 2010 to 2021, covering the academic years from 2010-11 to 2021-22, and includes the entire population of Unimib students, including those without migrant backgrounds. For each year, the dataset contains records (one for each enrolled student that year) ranging from 1,141 to 36,382.

From a statistical perspective, data from administrative archives, like other Big Data sources, can present quality issues. This has been widely discussed in the literature \citep{Baker2017} and generally stems from two main factors. On the one hand, typos, imprecise recording, and poor coding standards can result in missing values, data errors, and records with semi-structured or even unstructured characteristics. Indeed, the original Unimib administrative dataset required extensive pre-editing to correct data typos and standardize variable coding across records and years.
On the other hand, because they are routinely recorded for administrative purposes, data from archives, registers, and other digital sources often have limited suitability for addressing specific research questions and objectives. Although the Unimib administrative dataset provides comprehensive coverage and precise data,  after editing it offers approximately 30 workable variables, which primarily concern Unimib students' demographics, educational history, and university careers. However, it provides limited information about students' family and migrant backgrounds and generally lacks data closely related to the aims and objectives of UNI4ALL. This limitation of the available administrative data was the primary motivation for conducting an online survey specifically designed to collect original data on the migrant backgrounds of Unimib students, which serves as the second data source for this paper.

\subsection{Survey data}  
\label{subsec:SurveyData}

The online survey was conducted in March 2022, targeting  students enrolled at Unimib in the academic year 2021-22. This included both students who joined Unimib that year and those who had enrolled in previous years covered by the administrative dataset (2010-2021) and were still enrolled in the 2021-22 academic year.

Unlike the comprehensive administrative data described in the previous section, the survey generated sample data, providing only estimates due to its partial coverage. However, being specifically designed {\em ad hoc} for UNI4ALL, it offers sample data for a wide array of variables tailored to the research objectives, questions, and scope of the project. The survey dataset comprises 694 records (final sample size), each with data on 210 variables informed by  a questionnaire that included  circa 70 questions.

Table \ref{Tab:varinfo} provides details on the types of data and variables collected, categorized according to the 7 sections of the survey questionnaire.

\begin{table}[h!]
\caption{Types of data and number of variables per category from the two available sources}
\scriptsize
\centering
\begin{tabular}{lcc}
\toprule
& administrative & sample survey \\ 
\midrule
Discrimination &  & 7 \\
Family-Housing & & 25\\
Future research &  & 3  \\
Languages &  & 57 \\
School-University & 23 &  56 \\
Social networks  &  & 25  \\
Socio-Demographics & 6 & 32  \\
University services &  & 5 \\
\midrule
Total   &  29 & 210 \\
\bottomrule
\end{tabular}
\label{Tab:varinfo}
\end{table}

For example, the "Discrimination" section includes variables related to being discriminated during school or university due to the migratory background. In the "Family-Housing" section, we find variables such as  parents' education and the Italian region where the student is currently living. The "Future Research" section comprises the 3 variables indicating whether or not the student has suggestions concerning university services for students with migratory background, which service they would suggest and if they agree to be interviewed in a possible future wave of the same survey. The "Languages" section includes variables such as which languages does the student speak, what is the language they prefer to speak in specific situations and whether they can speak the same language of their parents. The "School-University" and "Socio-Demographics" sections, with 56 and 32 variables respectively, relate to student demographics, previous educational paths, and university careers at the time of the survey, e.g. the year of arrival in Italy and a self-assessment of their academic career up to the moment of the survey. Note that 2 "Socio-Demographics" variables and 8 "School-University"
variables are common with the administrative dataset, such as student gender, citizenship, secondary school type and university course of study.
Finally, the "Social Networks" and "University Services" sections group variables such as whether the student studies  with other colleagues on a regular basis, if they share the same nationality and whether or not they resort  to university services such as  tutorship or Italian language courses. It is important to note that the two datasets share data that allows for exact record linkage, enabling the integration of the (sample) survey data into the exact record in  the (complete) administrative dataset. This is a circumstance rarely addressed in the literature \citep{KimShao2022}, yet it arises in practice when the research commissioner also owns the complete dataset (proprietary database) and is authorized to use the data for statistical purposes according to the law (e.g. European GDPR), as is the case in this paper. Despite this, both the administrative and the survey data sets still present significant statistical challenges.

\FloatBarrier

\subsection{Statistical challenges } 
\label{subsec:SC}

To map Unimib students with a migrant background and analyze their characteristics for the UNI4ALL study, the available data poses challenges with both practical and methodological implications. The primary challenge is that the defined target population (see subsection \ref{subsec:TP}) is not completely identifiable in the complete administrative dataset. Significant portions of the target population are indistinguishable from students without a migrant background. This particularly affects students who meet only criterion 3 in our target population definition, concerning their parents' birthplace, which is not typically recorded in university archives. Consequently, the actual size of the study population is an unknown integer $N$, making it a migrant population parameter that needs to be estimated.
Moreover, for the purposes of the UNI4ALL online survey, this implies the absence of a sampling frame, i.e., a complete and updated list of the study population that would allow for the selection of a sample under a controlled sample design, with sample size $n$ chosen as a convenient fraction $n/N$ of the surveyed population. Operationally, the lack of a sampling frame has imposed sending an invitation email with a link to participate in the online survey to all Unimib students listed in the administrative archive. Over-coverage of the target population was managed through a two-step screening system. First, the invitation email included a message clarifying the survey objectives and the eligibility criteria (as listed in subsection \ref{subsec:TP}). Second, a screening question was administered on the opening page of the online questionnaire, to enforce the initial screening and exclude any accidental ineligible access.
This approach resulted in a self-selected opt-in sample, which has serious methodological implications for the quality of both the sample data and its statistical analysis. In survey sampling literature, this is known as a non-probabilistic sample, generated by an unknown, possibly unbalanced selection mechanism, and therefore unlikely to be representative of the study population \citep{BoonstraEtAl2021}. This lack of a probabilistic sample design means that the extent of representativeness cannot be statistically assessed. Consequently, na\"ive use of data from a non-probabilistic sample can lead to severely biased estimates and conclusions, unless appropriate statistical methodologies to reduce (and possibly correct) the selection bias are applied. Research in sampling methodology is increasingly focusing on this issue, which is connected to the widespread use of low-cost online surveys and to the new digital data sources broadly referred to as Big Data. Excellent reviews of such methodologies can be found, among others,  in \cite{Rao2021}, \cite{BeaumontRao2021} and \cite{ContiSIS2022}.
However, for the specific case study considered in this paper, completely mapping the target population, including the non-identifiable sub-groups discussed above, is a precondition for addressing the biased analysis of data from the online survey. In this paper, we contribute  an original method to identify, with measurable approximation, sub-groups of Unimib students that are member of the target population, and thus were  eligible for the UNI4ALL online survey,  but are statistically hidden within the entire list of Unimib students due to a lack of administrative data.
As a result of this first step, the administrative data set will be integrated with two new variables: “membership in the migrant population/eligibility for the survey” and “typology of migration backgrounds.” Both variables can be exactly assigned based on the available data only for a part of the target population. For hidden, eligible sub-groups, these variables will be imputed based on the methodology proposed here. The new variable “membership/eligibility” will be binary, while the new variable “typology of migration backgrounds” will be categorical.
The methodology proposed involves a multi-step process illustrated in the next Section.

As noted above, this first step is a prerequisite for addressing the second challenge: dealing with biased survey data from a non-probability (self-selected) sample, which will be tackled in further research. 

\section{Methodology}
\label{sec:method}

In this section, we illustrate the multi-step methodology proposed to fully identifying the target population. To achieve this, we treat the non-identifiable portion of the target population as a single subgroup of students who meet only criterion 3) in the definition in \ref{subsec:TP}, namely, Italian subjects born in Italy with at least one parent of foreign citizenship at birth. We also focus on the cross-section of the complete administrative dataset for the academic year 2021-22 to ensure comparability with the survey data. As the first step in the proposed methodology, we introduce a formal framework and appropriate notation.

\subsection{Modelling Target Population Membership  and Distinct Types of Migrant Background }
\label{subsec:TP&E}

Let $\cal U$ denote the target population of students with a migrant background. Let $i$ index each record in the cross-section 2021-22 of the administrative dataset, referring to all students enrolled at Unimib that academic year, regardless of their migration background ($i=1...36,\!382$ ). The (non-random) indicator
\begin{equation}
\label{Eq:delta}
\delta_i= \begin{cases}
1 & \text{if }i \in {\cal U} \\
0 & \text{otherwise}
\end{cases}
\end{equation}
defines membership in the target population and indicates eligibility for the online survey. Additionally, it allows for the formal definition of the unknown size of our target population as $\sum_i\delta_i=N$. This is a key parameter for the UNI4ALL study. However, $N$ is only partially known from the available data since $\delta_i$ is not fully observed in the administrative dataset.
To clarify this, we consider an equivalent, though operationally more convenient, expression for $\delta_i$ that is explicitly related both to the three criteria defining $\cal U$ and to the available data (see Section  \ref{sec:data}). To this end, we define three further indicators for all Unimib students $i$ included in the administrative dataset for the academic year 2021-22 :

\begin{equation}
BP_i= \begin{cases}
1 & \text{ if student $i$ was born in Italy}\\
0 & \text{otherwise}
\end{cases}
\end{equation}
which identifies students by their birthplace, indicating whether they meet (do not meet) membership/eligibility criterion 1);

\begin{equation}
CIT_i= \begin{cases}
1 & \text{if unit $i$ has Italian citizenship} \\
0 & \text{otherwise}
\end{cases}
\end{equation}
which identifies students by their citizenship, indicating whether they meet (do not meet) membership/eligibility criterion 2); and

\begin{equation}
PA_i= \begin{cases}
1 & \text{if both parents of unit $i$ had Italian nationality at birth} \\
0 & \text{otherwise}
\end{cases}
\end{equation}
which identifies students whose parents both had Italian nationality at birth, indicating whether they do not meet (meet) membership/eligibility criterion 3).

The operational definition of $\delta_i$ is then based on the product of the three indicators above and is equivalent to (\ref{Eq:delta}):
\begin{equation}
\label{Eq:delta2}
\delta_i=|BP_i \cdot CIT_i \cdot PA_i - 1|.
\end{equation}
Here, $\delta_i$ equals 1 if at least one of the indicators $BP_i$, $CIT_i$, or $PA_i$ equals 0. This is consistent with both Equation (\ref{Eq:delta}) and the complete set of criteria in subsection \ref{subsec:TP}, defining membership in $\cal U$ and survey eligibility. Equation (\ref{Eq:delta2}) also clarifies why the target population, and consequently its exact size $N$, are not completely known: the indicator $PA_i$ is not observed in the administrative dataset.

In principle, Equation \ref{Eq:delta2} gives rise to $2^3=8$ combinations of the indicators $BP$, $CIT$, and $PA$. However, not all combinations are relevant for our study. The combinations $0 \ 0 \ 1$, indicating a student with Italian parents who is born abroad and has foreign citizenship, and $1 \ 0 \ 1$, indicating a student with Italian parents who is born in Italy and has foreign citizenship, are excluded according to Italian law, which is based on the {\em Jus Sanguinis} principle. Additionally, an exception should be noted for a student $i$ with foreign birthplace  $(BP_i=0)$, Italian citizenship $(CIT_i=1)$, and Italian parents $(PA_i=1)$. This includes cases like a child born to Italian parents temporarily abroad. For UNI4ALL research purposes, this case is not included in the target population and is not eligible for the survey. 
Thus, for the combination $0 \ 1 \ 1$, Equation  (\ref{Eq:delta2}) is set to equal $0$. Such cases are expected to be limited in the entire administrative dataset, and indeed, a (statistically) negligible number of instances were identified in the survey data. Therefore, this case is treated as equivalent to any other student not included in the target population, indicated by the combination $1 \ 1 \ 1$. 

The remaining 5 relevant combinations of the indicators $BP_i$, $CIT_i$, and $PA_i$ in (\ref{Eq:delta2}) formally identify significant sub-groups within the target population, representing distinct types of migrant background. Specifically, the binary variable $\delta_i$: Target Population Membership/Survey Eligibility can be further detailed as the multilevel variable $\delta_i type$: Types of Migrant Background, which takes the following five values for each student $i$ in the administrative dataset:

\begin{equation}
\label{Eq:deltatype}  
\delta_i  type = \left\{
\begin{array}{cl}
0 & \text{ if } i \notin \cal U \text{ (no migrant background)} \\
1 & \text{ if }\text{Second generation with Italian citizenship} \\
2 & \text{ if }\text{Second generation with foreign citizenship} \\
3 & \text{ if }\text{Italian citizen with migrant experience} \\
4 & \text{ if }\text{Foreign citizen} \\
\end{array} \right.
\end{equation}

This is summarized in Table \ref{Tab:1}.

\begin{table}
 \caption{Relevant combinations of the indicators $BP$, $CIT$, $PA$ and definition of Membership/Eligibility $\delta_i$ and of types of migrant background $\delta_i type$ }
    \centering
    \begin{tabular}{cccccc}
\toprule
    \scriptsize Birth &  &  \scriptsize Parents' birth &  \scriptsize{Pop Member/} &  & \scriptsize Migration  \\
 \scriptsize place &  \scriptsize Citizenship &  \scriptsize nationality &  \scriptsize{Eligibility} &  & \scriptsize  background \\ \\
        $BP_i$ & $CIT_i$ & $PA_i$ & $\delta_i$ &     $\quad \delta_i type$ & \\
    \midrule
         1 & 1 & 1 & 0 &  0 & \scriptsize{no migrant} \\    
         & & & & & \scriptsize background\\ \\
         1 & 1 & 0 & 1 &  1 &  \scriptsize{$2^{nd}$ generation} \\
        & & & & & \scriptsize{Italian } \\ \\
        1 & 0 & 0 & 1 & 2 & \scriptsize{$2^{nd}$ generation} \\ 
         & & & & & \scriptsize{ foreign} \\ \\
        0 & 1 & 0 & 1 &  3 & \scriptsize{Italian with} \\ 
        & &  &  &  & \scriptsize{migrant experience}\\ \\
        0 & 0 & 0 & 1 & 4 & \scriptsize{Foreign} \\    
\bottomrule
    \end{tabular}
    \label{Tab:1}
    \vspace{0.5 cm}
\end{table}

Being able to identify our target population in the complete administrative dataset would allow the integration of both variables 
$\delta_i$ and $\delta_i type$ into it for all records $i$. This is a main aim of this paper: it would enable the expansion of the official database of all Unimib students with relevant demographics concerning their migrant status and history, thereby allowing the analysis of these strata of Unimib students with respect to all variables registered in the administrative dataset. However, it is crucial to note that the administrative dataset provides complete and accurate information on students' birthplace and citizenship, but no migrant-related information regarding their parents. Parents' birth nationality is only available for students included in the survey sample.
As a result, both $\delta_i$ and $\delta_i type$ can be precisely determined for all records $i$ limited to three subgroups of the target population (Table \ref{Tab:1} rows 3 to 5). In contrast, the important segment of the population comprising second-generation Italian students remains indistinguishable from students with no migrant background in the complete dataset (Table \ref{Tab:1} rows 1 and 2).

In terms of formal indicators, computing both $\delta_i$ in (\ref{Eq:delta2}) and $\delta_i type$ in (\ref{Eq:deltatype}) makes the value of $PA_i$ redundant, except when distinguishing students with no migrant background from second-generation Italian students (1\textsuperscript{st} and 2\textsuperscript{nd} rows in Table \ref{Tab:1}). For this latter distinction, $PA_i$ is actually necessary and effective.
Thus, the second step in our methodology involves linking survey data with administrative data based on the available exact linkage key, as discussed in \ref{subsec:SurveyData}. This allowed for the precise identification of second-generation Italian students (Table \ref{Tab:1} row 2) limited to those included in the sample. To progress further toward our goal of fully integrating $\delta_i$ and $\delta_i type$ into the administrative dataset, the final step of the proposed methodology involves leveraging all available data to {\em predict} (or {\em impute}) $PA_i$ for all students $i$ who did not participate in the survey. This will also provide estimates of the size of the study population $N$ and the sizes of the main sub-populations with different types of migration backgrounds, as outlined in Table \ref{Tab:1}.

\subsection{Predicting  Membership in the Second Generation Italian Subgroup}
\label{subsec:Prediction}
Two prediction  approaches, a classical statistical parametric model and a popular machine learning technique, are considered and compared to identify all students with  migrant backgrounds in the complete administrative dataset. In this final step, we focus on the combinations $1 \ 1 \ 1$ and $1 \ 1 \ 0$ from the first two rows of Table \ref{Tab:1}. Specifically, we are interested in units $i$ where both $BP_i$ and $CIT_i$ are equal to 1. This subgroup of Unimib students is the only one where the value of $PA_i$ influences $\delta_i$ and $\delta_i type$. For all other units $i$, Table \ref{Tab:1} shows that $PA_i$ is consistently 0, resulting in $\delta_i=1$ regardless of the migrant background type and the value of $\delta_i type$. Therefore, we concentrate on students $i$ with $BP_i=CIT_i=1$ and discuss the data requirements for predicting $PA_i$. 

The prediction model aims to classify which subjects $i$ in the focused group of Italian students born in Italy are included in $\cal U$ for having at least one parent with foreign citizenship at birth ($PA_i=0$), and which are not ($PA_i=1$). Both approaches described here require a sufficiently large dataset with information on parents' citizenship at birth. Since $PA_i$ is only observed in the online survey, i.e., whether $i$ is included in the self-selected sample, no usable data is available for all non-eligible students, either in the complete administrative archive or the survey dataset. To address this, we leveraged the fact that the first phase of the online survey's screening system was unsuccessful in a significant number of cases. We accessed metadata for 402 non-eligible students stored in the survey platform as cases excluded from the survey based on the second-phase screening question. We combined these with eligible students who answered  the survey and have $BP_i=CIT_i=1$, since, as already mentioned, this is the only subgroup where the value of $PA_i$ influences $\delta_i$ and $\delta_i type$. The selection of this subgroup reduces the survey sample size from 694 to 312 students. A new data set comprising 714 students ($402$ with $PA_i=1$ and $312$ with $PA_i=0$) that includes variables common to both the administrative and survey datasets is thus formed for prediction purposes. The final set of predictors was empirically selected based on their effectiveness in predicting the outcome $PA_i$. Cross-correlation was also checked to avoid multicollinearity.  \autoref{Tab:catvar} and \autoref{Tab:quantvar} list the selected predictors of mixed nature,  which include student's gender, course of enrollment, employment status, and the number of ECTS (Italian university CFU) earned at the time of the survey. Additionally, a binary (dummy)  variable indicating whether the student's given name is among the most common in Italy was included. This variable is based on an external, publicly available data source, namely a large database of more than 32,000 given names of Italian citizens (Anagrafe amministratori locali 1985-2014, \url{ https://dait.interno.gov.it/elezioni/anagrafe-amministratori}). The rationale of using this further predictor is that the commonness or rarity of a student's first name empirically proved to provide evidence regarding the nationality at birth of the parents, thus enhancing the predictive power of our models. The same set of predictors was used in both prediction approaches.

\begin{table}[ht]
\centering
\caption{Selected Categorical Predictors and Frequency Distribution in the dataset of size 714 for prediction purposes}
\begin{tabular}{@{}lll@{}}
\toprule
Variable & Levels & Frequency (\%) \\ 
\midrule
Employment & worker-student ( $<$50\% of study time) & 10.22\% \\ 
Employment & student-worker (50-75\% of study time) & 17.67\% \\ 
Employment & student ( $>$75\% of study time) & 71.55\% \\ 
Employment & not available & 0.56\% \\ 
Course of study & bachelor & 69.47\% \\
Course of study & master & 15.41\% \\
Course of study & bachelor \& master & 15.12\% \\
Gender & Female & 75.77\% \\
Gender & Male & 24.23\% \\
Common Italian name & No & 26.89\% \\
Common Italian name & Yes & 73.11\% \\
$PA_i$ & 0 & 43.70\% \\
$PA_i$ & 1 & 56.30\% \\
\bottomrule
\end{tabular}
\label{Tab:catvar}
\end{table}

\vspace{0.5cm}

\begin{table}[ht]
\centering
\caption{Selected Quantitative Predictors and Elementary Statistics}
\begin{tabular}{@{}lcc@{}}
\toprule
Statistic & years of enrollement at 2021/22 & earned ECTS at 2021-22 \\ 
\midrule
Mean & 2.838 & 31.483 \\ 
Std Deviation & 2.577 & 21.082 \\ 
Median & 2 & 0 \\ 
Mode & 1 & 32 \\ 
Min & 1 & 0 \\ 
Max & 24 & 113 \\ 
\bottomrule
\end{tabular}
\label{Tab:quantvar}
\end{table}

\vspace{0.5cm}
The classical approach uses a multivariate logistic regression to  model the probability
\begin{equation}
\log\left\{\frac{P(PA_i=0)}{1-P(PA_i=0)}\right\} = \log\left\{\frac{P(PA_i=0)}{P(PA_i=1)}\right\} 
\end{equation}
as a linear combination of all predictors in \autoref{Tab:catvar}  and \autoref{Tab:quantvar}  as independent regressors. The logistic regression has been implemented employing the \textit{Stats} base package in R Statistical Software \citep{RCoreTeam2021}.

The machine learning approach uses a random forest classifier.  Automated model-selection,  the ability to detect higher order interactions and complex relationships, demonstrated  remarkable predictive performance  make this choice appealing  among other machine learning tools (see for instance \cite{GenuerPoggi2020}). The random forest classifier has been implemented employing the \textit{randomForest} package \citep{Liaw2002} in R Statistical Software \citep{RCoreTeam2021}.

Both models were validated for their ability to predict the outcome $PA_i$ for all students $i$ in the focused subgroup. The validation procedure involved splitting the available data into training and validation subsets in a $0.75/0.25$ ratio. Predictive ability was measured on the validation subset using the following metrics implemented in the \textit{caret} package \citep{Kuhn2008} in R Statistical Software: 

\begin{equation}
  \text{Accuracy} = \dfrac{\text{True Positives} + \text{True Negatives}}{\text{Total Instances}} 
\label{eq:Accuracy}
\end{equation}

\begin{equation}
\text{Precision} = \dfrac{\text{True Positives}}{\text{True Positives} + \text{False Positives}}
\end{equation}

Focusing on false negatives 
\begin{equation}
\text{True positive rate} = \dfrac{\text{True Positives}}{\text{True Positives} + \text{False Negatives}}
\label{Eq:rec}
\end{equation}

and by synthesing both True positive rate and Precision
\begin{equation}
\text{F1 Score} = \dfrac{2 \times \text{Precision} \times \text{True positive rate}}{\text{Precision} + \text{Recall}}.
\end{equation}

Table \ref{tab:model_comparison} presents the metrics for both models. The high levels of accuracy ( $\ge 70\%$) and precision ($\ge 80\%$) are particularly noteworthy, as they are uncommon in machine learning applications to real-world data. This highlights the models' good  predictive capabilities, with logistic regression performing slightly better.

\vspace{0.5cm}
\begin{table}[ht]
\centering
\caption{Performance metrics for the two predictive models employed} 
\begin{tabular}{ccccc}
\toprule
{ Model} & {Accuracy     } & {Precision } & {True positive rate} & {F1 Score }  \\ 
\midrule
Logistic Regression & 0.76 & 0.83 & 0.60 & 0.70 \\ 
  Random Forest & 0.74 & 0.80 & 0.59 & 0.68 \\ 
\botrule
\end{tabular}
\label{tab:model_comparison}
\end{table}

\begin{figure}[t] \centering \includegraphics[width=0.8\linewidth]{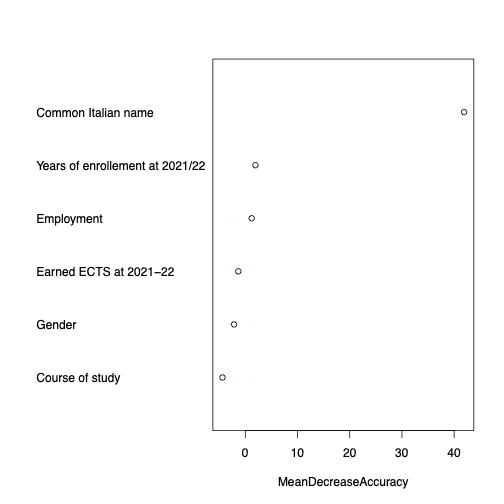} \caption{Variable Importance Plot for the fitted Random Forest model} 
\label{fig:VarImpPlot} 
\end{figure}

The sensitivity of the logistic (best performer) model to the chosen size of the training and validation sets, has been tested via a $k$-fold cross-validation procedure $(k=10)$, implemented  through the \textit{caret} package \citep{Kuhn2008} in R Statistical Software. Results do not highlight significant sensitivity of the logistic model to different choices of the size of the training and validation sets. Additionally, Cohen's Kappa index was also computed (by \textit{caret} R function) to check the robustness of the both models given the uneven distribution of the outcome $PA$. The result, 0.45 ($< 0.5$), confirms that the models' performance is not merely due to chance.
Two additional visualization diagnostics have been considered: a variable importance plot for the random forest model (Breiman, 2001) and the ROC curve (McNeil \& Hanley, 1984) for the logistic regression model.
The variable importance plot illustrates the influence of each predictor on the overall predictive power of the fitted model. It is based on the relative contribution of each predictor to the model's accuracy (\ref{eq:Accuracy}), calculated as the Mean Decrease in Accuracy (MDA) when each predictor is excluded one at a time. For our set of predictors (see \autoref{Tab:catvar}
 and \autoref{Tab:quantvar}), the variable importance plot is shown in Figure \ref{fig:VarImpPlot}.

Key predictors identified by the analysis include having a common Italian name and years of enrollment, which show the highest MDA values. These variables are crucial to the model, as their exclusion significantly reduces the model's accuracy. Conversely, variables such as the course of study have lower MDA values, indicating that while they contribute to the model's performance, their impact is relatively minor. This analysis highlights the importance of specific academic and demographic factors in predicting the outcome $PA_i$, with having a common Italian name being the most influential predictor.

\begin{figure} [t]
\centering \includegraphics[width=0.8\linewidth]{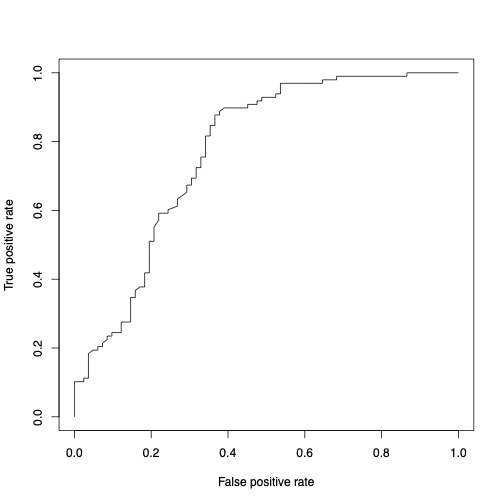}
\caption{ ROC curve for the fitted Logistic model} \label{fig:ROC} 
\end{figure}
Finally, the ROC (Receiver Operating Characteristic) curve provides a visualization of the logistic regression model's predictive performance. The curve plots the True positive rate as defined in Equation \ref{Eq:rec}, against the False positive rate, defined as $\textit{False positive}/\left({\textit{False positive} + \textit{True negative}}\right)$, across the possible values of the decision criterion \citep{swets1979}. In our case of logistic regression, the decision criterion governing the prediction procedure is the model predicted probability. Therefore the trade-off between True positive rate and False positive rate  is evaluated on the probability space, across several thresholds in  $(0,\,1)$. A wider area under the curve (AUC) indicates better model performance, with a curve approaching the top left corner of the graph, signifying high True positive rate and low False positive rate.
The ROC curve for our fitted logistic model is shown in \autoref{fig:ROC}. It reveals an AUC of 0.773, indicating a fair level of discriminatory power (77.3\%) that is significantly above average. This result supports the model's effectiveness in balancing the ability to correctly identify true positives and true negatives, making it a reliable tool for predictive analysis.

\section{Results and Discussion}
\label{sec:resdisc}
Adding the complete administrative archive with the new variables $\delta_i$ and $\delta_i type$, which fully identify the study population, allows for its statistical analysis, starting with its total size ($N$) and the size of its relevant subgroups, as shown in Table \ref{Tab:Res}. Note that the distributions in Table \ref{Tab:Res} are predicted (estimated) based on the (best performer) Logistic model in Section \ref{subsec:Prediction} in the second row, while they are computed exactly for all other cases.

Moreover, the distributions in Table \ref{Tab:Res} align with expert opinions as well as with other studies on  university students with migrant backgrounds. Additionally, these  distributions  appear consistent with  recent data published on second generations \citep{Conti2020} and on foreign students \citep{Giudici2021,Aiello2020} although these sources, particularly those focusing on university students, consider only one typology among those defined in \autoref{Tab:Res}, namely international students who have both a foreign place of birth and citizenship. It is notable that the weight of this component (slightly more than 5\%) is similar to findings at Unimib, is similar to findings at Unimib, keeping in mind that the analyses published so far for Italy, to the best of the authors' knowledge, distinguish within this typology between those who obtained their high school diploma abroad and those who obtained it in Italy, identifying the latter group as the second generation.

The integration of both $\delta_i$ and $\delta_i type$ also allows for the analysis of all variables recorded in the complete administrative dataset, specifically for the study population and its relevant subgroups. This includes, for instance, the study of university performance among different components of the student population with a migrant background in comparison to native students (e.g. drop-outs, current graduates, graduation grades, CFU acquired), as well as  analyses of the support networks (e.g. friendships) and the use of services specifically arranged by the university to facilitate international students and those with a migrant background.

\begingroup
\setlength{\tabcolsep}{4pt}
\begin{table}[ht]
\caption{Distribution of $\delta_i$ and $\delta_i type$ integrated in the Unimib admin. dataset}
\centering
\scriptsize
\begin{tabular}{ccrrrc}
  \hline
  & &  & $\%$ (TOT$=36,382$ & $\%$ (TOT$=5,492$  & \\
  $\delta_i$ & $\delta_i type$&  Freq.    &  All Unimib stud. & stud. with a & Migrant backgrounds\\
   &                &        &   2021-22) &   migrant background) \\
  \hline
     0 & 0 & 30,891 & 84.91 & & No migrant background \\ 
      1&1& 2,828& 7.77& 51.50 & 2nd generation Italian\\ 
         1&2& 132& 0.36& 2.40 &  2nd generation foreign\\ 
         1&3& 662& 1.82& 12.06 & Italian with migrant experience \\ 
 1&4&1,869& 5.14 & 34,04 &  Foreign \\ 
   \hline
\end{tabular}
\label{Tab:Res}
\end{table}
\endgroup

\begin{figure}[t]
   \centering
   \includegraphics[trim=0 16 0 0, clip, width=0.8\linewidth]{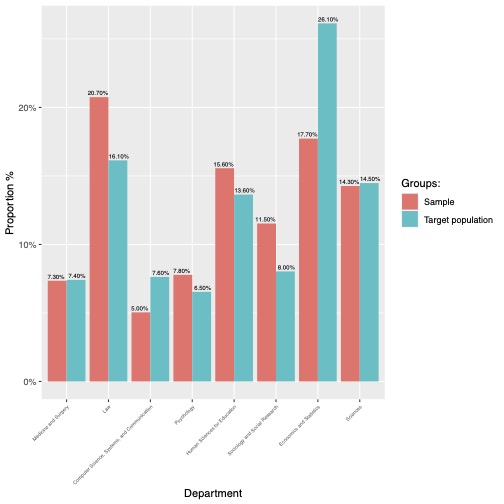}
       \caption{Distribution of the typology of academic Course (Departments) in the sample and in the target population}
   \label{fig:Department}
\end{figure}

\begin{figure}[t]
    \centering
    \includegraphics[trim=0 16 0 0, clip,width=0.6\linewidth]{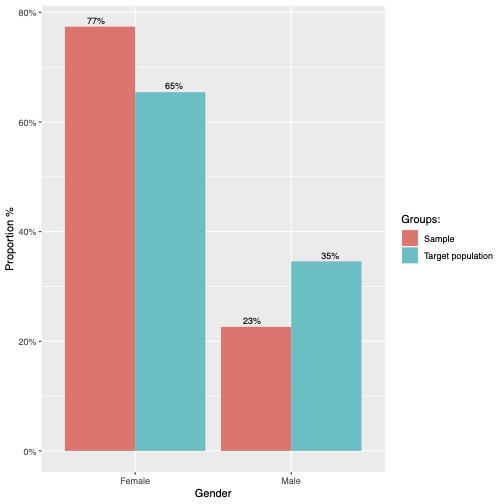}
        \caption{Distribution of gender in the sample and in the target population}
    \label{fig:gender}
\end{figure}

\begin{figure}[t]
    \centering
    \includegraphics[trim=0 16 0 0, clip,width=0.6\linewidth]{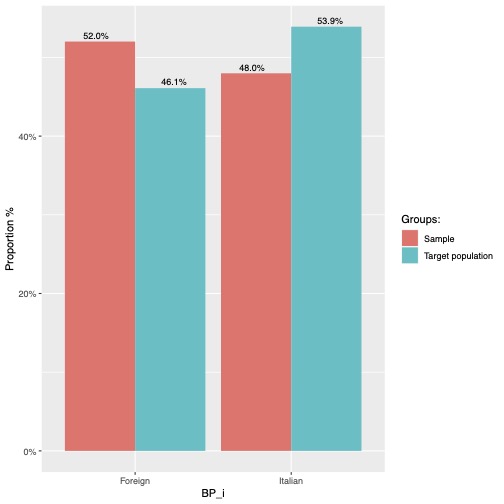}
          \caption{Distribution of birth place in the sample and in the target population}
    \label{fig:Ibp}
\end{figure}

\begin{figure}[t]
    \centering
    \includegraphics[trim=0 16 0 0, clip,width=0.6\linewidth]{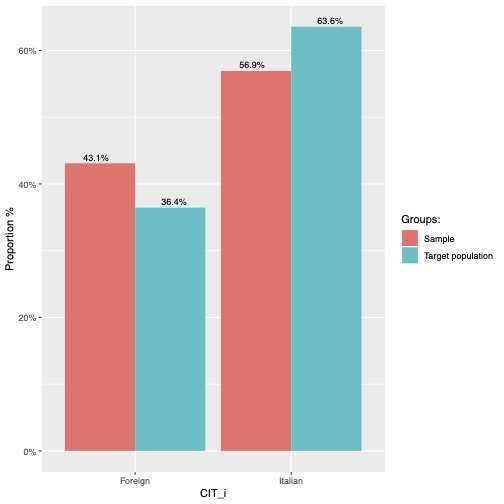}
        \caption{Distribution of citizenship in the sample and in the target population}
    \label{fig:Icit}
\end{figure}

Finally, identifying  our target population as defined in Section \ref{subsec:TP}, allows for a broad assessment of the potential bias in the original data collected through the self-selected online survey. This issue has been  extensively discussed in Section \ref{subsec:SC} as a primary challenge due to the lack of a probabilistic structure in the survey  sample. The assessment focuses on variables  common to both  the complete Unimib administrative dataset and  the UNI4ALL survey sample dataset,  comparing the true target population proportions with the corresponding   proportions computed from  the  survey sample data. 

Figures \ref{fig:Department} to \ref{fig:Icit} provide  examples of such comparisons,  highlighting disproportions and imbalances in the sample data across 4 variables. For instance, we observe notable over- and under-representations in the sample (up to 8.4 percentage points) for students' choice of academic study type (Figure \ref{fig:Department}) as well as substantial biases across the  additional  3 variables that are even more critical to the the UNI4ALL project's scope and objectives. Specifically, the sample shows a 12-percentage-point  under-representation of male students, and biases of 5.9 and 6.7 percentage points, respectively, in students' birth place and citizenship at the time of the survey, whether foreign or Italian.   

Although this broad assessment is limited to a few variables, each analyzed individually, it clearly indicates a significant selection bias in the opt-in survey sample, preventing the straightforward use of this data to produce accurate and reliable estimates.

Naturally, the methodology proposed in this paper has limitations. For instance, in Section \ref{sec:method}, our choice to use and compare logistic regression and random forest models is somewhat arbitrary; other statistical and machine learning methods could have been used for prediction, potentially yielding different results and levels of accuracy. However, we believe that the most crucial factor for substantially increasing predictive accuracy in this application is the quality and quantity of predictors. The results presented here are constrained by the limited number of variables with strong predictive power for migrant background available in the Unimib administrative dataset. This limitation led us to seek external data sources, which is not always feasible due to quality and confidentiality issues. The collection of richer and more detailed administrative data should therefore be encouraged and prioritized.

On the other hand, this work demonstrates potential for guiding future surveys within universities and other organizations, particularly when focusing on specific population strata and subgroups that may be partially hidden or not fully identifiable through routinely registered internal datasets for administrative or other reasons. In this regard, the Unimib UNI4ALL experience considered here offers some practical insights. An inexpensive and timely approach using a non-probability online sample of volunteers can be effective, provided that a well-designed screening process is implemented. According to the strategy proposed in Section \ref{sec:method},  this screening should collect a substantial amount of supplementary information to facilitate the application of out-of-sample statistical and machine learning techniques for predicting membership in the population group of interest. In fact, if the Unimib survey had employed a single-step, broad screening—allowing all students (with or without a migrant background) to access the online survey—it would have gathered a larger dataset for predicting membership in the hidden subpopulation, thereby improving prediction accuracy.

In conclusion, the comprehensive identification of the Uni4All target population, as presented in this paper, was a necessary and essential step toward developing suitable bias-reducing methodologies for the UNI4ALL survey data, which will be the focus of future research."
\FloatBarrier

\bibliography{bibliography}

\end{document}